\documentclass{article}
\usepackage{graphicx}
\usepackage{listings}
\usepackage{amsmath}
\usepackage{hyperref}
\usepackage{enumitem}
\usepackage{float}
\usepackage[sorting=none]{biblatex}
\lstdefinestyle{mystyle}{
    language=Python, 
    basicstyle=\ttfamily\small, 
    frame=bt, 
    columns=fullflexible,
    aboveskip=10pt,
    belowskip=10pt,
    tabsize=4,
    keepspaces=true,
    showstringspaces=false
}

\title{Confidential Wrapped Ethereum}
\author{Artem Chystiakov, Mariia Zhvanko \\ Distributed Lab}
\date{June 2025}

\begin{document}

\maketitle

\section{Introduction}
Transparency is one of the key benefits of public blockchains. However, the public visibility of transactions potentially compromises users’ privacy. The fundamental challenge is to balance the intrinsic benefits of blockchain openness with the vital need for individual confidentiality \cite{Cav2025}.
\par The goal is to create a permissionless, public good protocol that will obfuscate users’ financial activity within ETH tokens by encrypting their balances and transfer amounts. This will allow for confidential peer-to-peer payments, donations, and acquisitions in ETH without relying on centralized entities.
\par The proposal suggests creating a confidential version of wrapped Ethereum (cWETH) fully within the application layer. The solution combines the Elliptic Curve (EC) Twisted ElGamal-based commitment scheme to preserve confidentiality and the EC Diffie-Hellman (DH) protocol to introduce accessibility limited by the commitment scheme. To enforce the correct generation of commitments, encryption, and decryption, zk-SNARKs are utilized. 

\subsection{Difference from the existing protocols} 
There are at least two known solutions that do not require the protocol layer modifications and can be implemented as is.
\par The Solana \cite{Sol22} and Zether \cite{BA20} approaches may seem very similar at first glance, as they also use ElGamal commitments. However, the main distinguishing factor is the need to solve the discrete logarithm problem to access the encrypted balance. 
\par Solana partially mitigates this by maintaining a separate decryptable balance. But the main difference is that the encryption scheme used doesn’t support aggregation, and such a balance works mainly as a cache, which is updated when the pending amount is moved to the actual one. However, it is still required to decrypt the balance represented as the ElGamal commitment, which reintroduces the necessity to solve the discrete logarithm problem. Although this process can be simplified by dividing the value into chunks, the cWETH protocol proposes an encryption scheme that is aggregateable and is managed along with the ElGamal commitment to avoid computing the discrete logarithm at all.
\par Another difference lies in the ZK proofs used by cWETH. Both Solana and Zether approaches rely on the Sigma protocols and bulletproofs, while this proposal is based on the zk-SNARKs. The tradeoff here is the need for a trusted setup, but this can be mitigated by using existing trusted setups, such as Plonk’s universal setup, which has proven to be secure over time.

\section{cWETH Overview}
\subsection{Wrapping ETH} 
The process of setting up the confidential account is considered to be a part of the ETH wrapping operation. Along with the usual wrap logic, upon the first deposit to the cWETH contract, each user provides a babyJubJub public key that will be used for the management of balances.
\par This public key is employed in the commitment and encryption of token amounts and balance computations. The key pair generation and initial deposit flow is depicted in the following diagram:

\begin{figure}[H]
    \centering
    \includegraphics[width=\textwidth]{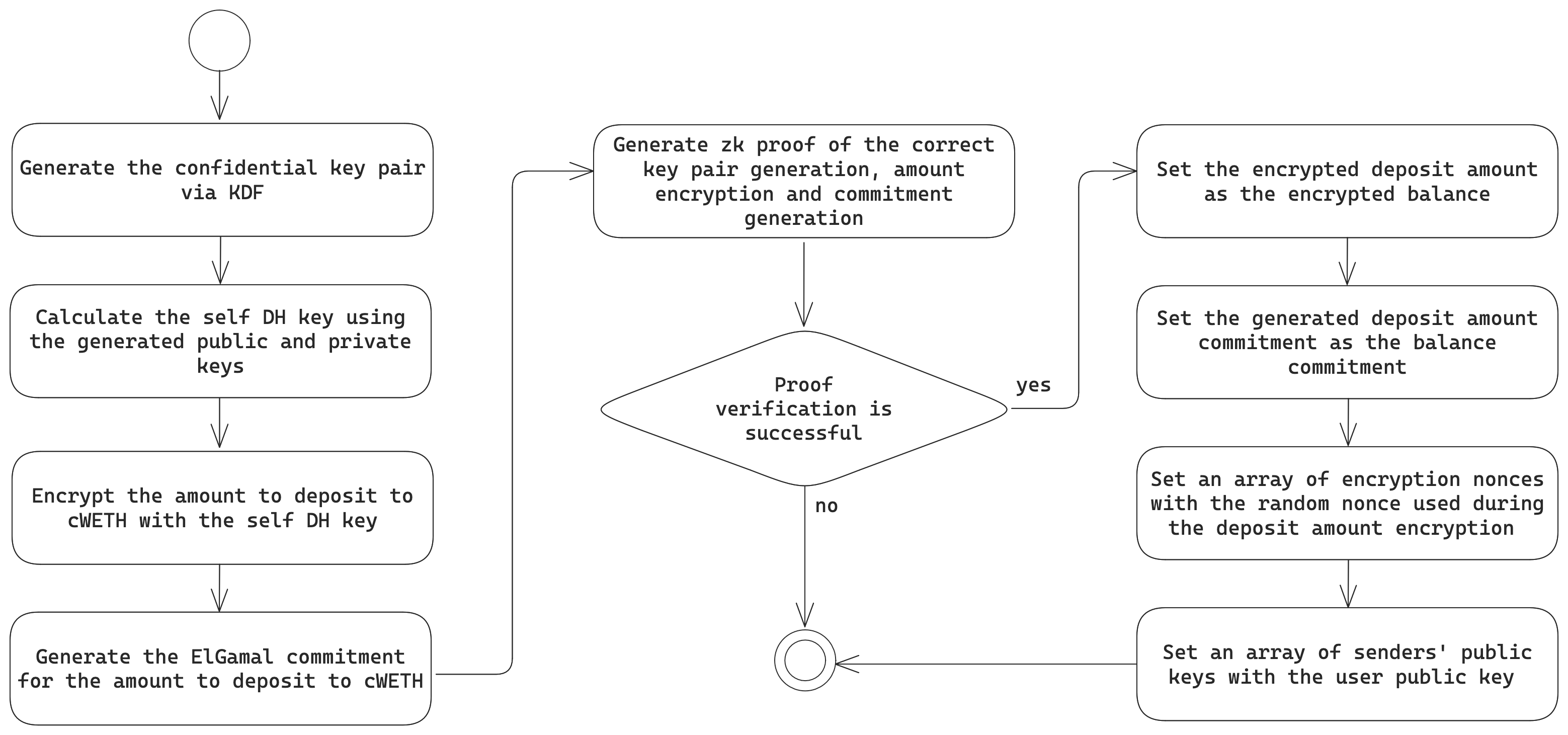}
    \caption{Initial deposit to cWETH flow}
    \label{fig:mesh1}
\end{figure}

\par After the first deposit to the cWETH contract, initial balances are updated with the values calculated according to the same formulas used for a confidential transfer, considering that the user is both the sender and the receiver. Technical details related to these values and the key pair generation process are provided in further sections of this proposal.
\par The unwrapping operation is very similar conceptually, hence it will not be covered in this section.

\subsection{Balance management} 
To avoid computing the discrete logarithm to calculate the balances hidden with ElGamal, two parallel representations of balances are maintained:

\begin{itemize}
  \item \textit{ElGamal commitment} that is only needed for ZK proofs verification of the balance knowledge.
  \item \textit{Balance encrypted with the DH shared key} which enables users to decrypt their actual balance data. Access to the senders’ public keys array and the encryption nonces array is required to perform the decryption.
\end{itemize}

\par Because ZK proofs are generated from the user's current balance, changing it before the transaction execution may invalidate the proof. To address this issue, both types of balance (ElGamal commitment and DH-encrypted) have to be presented in two separate states:

\begin{itemize}
  \item Pending balance to receive tokens.
  \item Actual balance to spend tokens.
\end{itemize}

\par Users can move tokens from their pending balance to the actual one at any time. Additionally, it could be done at the end of each transfer initiated by the user. This approach will consolidate the balance update into a single operation, reducing the gas usage compared to two separate transactions.

\subsection{Confidential transfer}
To initiate a confidential transfer, it is necessary to provide the token amount represented in four different forms:

\begin{itemize}
  \item Hidden inside the receiver's public key-based ElGamal commitment.
  \item Hidden inside the sender's public key-based ElGamal commitment.
  \item Encrypted with the receiver's public key-based DH shared key for incrementing the receiver’s balance.
  \item Encrypted with the sender’s public key-based DH shared key for the new sender’s balance.
\end{itemize}

The diagram below illustrates the confidential transfer process:

\begin{figure}[H]
    \centering
    \includegraphics[width=\textwidth]{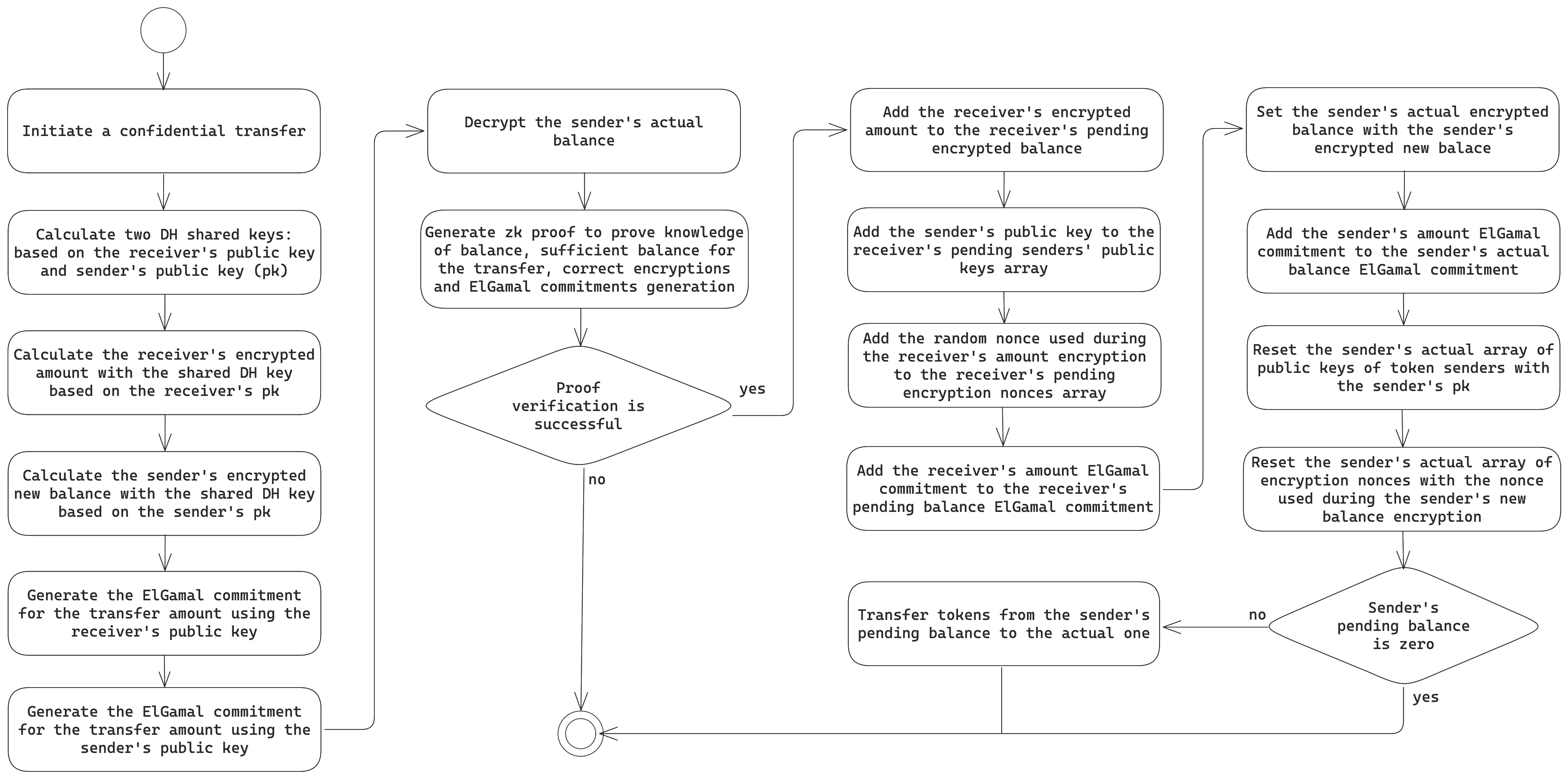}
    \caption{Confidential transfer flow}
    \label{fig:mesh2}
\end{figure}

\par  During the transfer, both types of balances and their associated amounts are updated to ensure consistency between different balance representations.
\par Note that there exist two separate versions of the same values for both encrypted with the DH shared key and ElGamal commitment representations: the sender’s public key-based and the receiver’s public key-based. This is caused by different approaches used during the encryption and commitment generation, technical details of which are provided in the further sections of the proposal.

\section{Confidentiality Math}
\subsection{KDF for confidential key pair} 

The protocol utilizes the deterministic KDF approach to generate a confidential key pair. The user has to sign an EIP-712 structured message with their Ethereum (ECDSA) private key. The hash of the signature is the babyJubJub private key.
\par The EIP-712 message is obtained as follows:

\begin{lstlisting}
bytes32 KDF_MSG_TYPEHASH = keccak256("KDF(address cWETHAddress)");

bytes32 kdfStructHash = keccak256(
    abi.encode(KDF_MSG_TYPEHASH, cWETHAddress)
);
\end{lstlisting}

\par The babyJubJub private key is obtained as follows:

\begin{lstlisting}
signature = eth_signTypedData_v4(kdfStructHash);
privateKey = keccak256(keccak256(signature));
\end{lstlisting}

\par Note that it is crucial to never reveal the signature as the private key is directly derived from it.
\par The key pair must satisfy the following:

\[ P = sk \times G, \]

\par where $P$ --- public key,
\par \hspace*{25px} $sk$ --- private key,
\par \hspace*{25px} $G$ --- base point on the elliptic curve.

\subsection{EC ElGamal-based commitment scheme}

The commitment based on the EC ElGamal scheme allows for a succinct representation suitable for ZK proofs. It also facilitates seamless balance management via its additive homomorphic properties.
\par The commitment of the balance consists of two parts and is constructed according to the formula:

\[C = b \times H + r \times G\]
\[D = r \times P\]

\par where $H$, $G$ --- base points on the elliptic curve,
\par \hspace*{25px} $b$ --- user balance,
\par \hspace*{25px} $r$ --- random nonce,
\par \hspace*{25px} $P$ --- user's public key.

\vspace{12pt}
\par To efficiently prove the commitment of the balance through a ZK proof, the user must use the private key to compensate for the lack of knowledge of the aggregated random nonce:

\[C = b \times H + sk^{-1} \times D = b \times H + r \times G\]

\par The commitment of the transfer amount is calculated differently for the sender and the receiver. The receiver’s amount commitment is derived as follows:

\[C_{a_r} = a \times H + r \times G\]
\[D_{a_r} = r \times P_r\]

\par where $P_r$ --- receiver’s public key.

\vspace{12pt}
\par The transfer amount commitment used in the aggregation of the sender’s balance commitment is calculated as:

\[C_{a_s} = r \times G - a \times H\]
\[D_{a_s} = r \times P_s\]

\par where $P_s$ --- sender’s public key.

\vspace{12pt}
\par The balance commitment is computed additively as outlined below:

\[C=\sum_{i=1}^N C_{a_i} = \sum_{i=1}^N a_i \times H + \sum_{i=1}^N r_i \times G\]
\[D=\sum_{i=1}^N D_{a_i} = \sum_{i=1}^N r_i \times P\]

\par where $N$ --- number of token transfers the user received,
\par \hspace*{25px} $C_{a_i}$, $D_{a_i}$ --- transfer amount commitment,
\par \hspace*{25px} $a_i$ --- amount transferred to the user,
\par \hspace*{25px} $r_i$ --- random nonce used in the commitment.

\subsection{Encryption with the EC DH shared key}
Since ElGamal commitments alone do not provide a convenient means for users to access decrypted balances, requiring solving the discrete logarithm problem, an additional form of amount hiding through encryption is employed using the EC DH shared keys.
\par The DH shared key is derived as follows:

\[K = sk_r \times P_s = sk_s \times P_r = sk_r \times sk_s \times G\]

\par where $sk_s$, $sk_r$ --- private keys of the sender and receiver, respectively,
\par \hspace*{25px} $P_s$, $P_r$ --- public keys of the sender and receiver, respectively.

\vspace{12pt}
\par The transfer amount DH-based encryption is performed differently for the sender and the receiver. 
\par The encryption of the transfer amount used to aggregate the receiver’s balance is calculated according to the formula:

\[A_r = a + K_x + poseidon(K_x || n)\mod p =\]
\[= a + (sk_s \times P_r)_x + poseidon((sk_s \times P_r)_x || n)\mod p\]

\par where $a$ --- amount to be transferred,
\par \hspace*{25px} $K_x$ --- $x$ coordinate of the DH shared key,
\par \hspace*{25px} $n$ --- random nonce,
\par \hspace*{25px} $p$ --- large prime number.

\vspace{12pt}
\par The random nonce usage is required to solve the problem of the potential transfer data leaks caused by the lack of randomness in the encryption scheme, which can be particularly noticeable in recurring payments. These nonce values have to be stored together with the senders’ public keys so that users can decrypt their balances.
\par Encrypted amounts are further used to aggregate the encrypted receiver’s balance:

\[B_r = \sum_{i=1}^N A_i \mod p =\]
\[= \sum_{i=1}^N a_i + \sum_{i=1}^N (sk_r \times P_{s_i})_x + \sum_{i=1}^N poseidon((sk_r \times P_{s_i})_x || n_i) \mod p\]

\par where $A_i$ --- amount encrypted with the DH shared key sent to the user.

\vspace{12pt}
\par According to the formula, the receiver can decrypt their balance as follows:

\[b_r = B_r - \sum_{i=1}^N (sk_r \times P_{s_i})_x \mod p - \sum_{i=1}^N poseidon((sk_r \times P_{s_i})_x || n_i) \mod p\]

\par On the sender side, using this approach for encrypting the transfer amounts could potentially lead to managing an infinite number of public keys and random nonces necessary for the balance decryption. To resolve this, each time a user transfers tokens, the lists of existing senders' public keys and random nonces are reset. This is accomplished by calculating the new encrypted sender's balance, as outlined below:

\[B_{s_{new}} = b_s - a + (sk \times P)_x + poseidon((sk \times P)_x || n) \mod p\]

\par where $n$ --- random nonce used in the new balance encryption,
\par \hspace*{25px} $b_s$ --- sender’s balance,
\par \hspace*{25px} $a$ --- amount to be transferred,
\par \hspace*{25px} $sk$ --- sender’s private key,
\par \hspace*{25px} $P$ --- sender’s public key.

\vspace{12pt}
\par After updating the encrypted balance, the sender needs to manage only their own public key and random nonce as a single entry to decrypt the total amount of tokens owned.
\par The identical logic for the new balance calculation can be used to unwrap cWETH tokens.

\section{Solidity Smart Contracts}
\subsection{State variables}
\subsubsection{Public keys}

First of all, each user is associated with the corresponding public key generated during the first deposit to the cWETH contract. This is managed in a simple ${mapping(address => uint256[2])}$.

\subsubsection{Balance}
Balance of each user is represented in the ElGamal commitment form:

\begin{lstlisting}
struct Commitment {
	uint256[2] C;
    uint256[2] D;
}
\end{lstlisting}

\par To decrypt the balance encrypted with the DH shared key, along with the DH balance itself, an array of public keys of the senders and an array of encryption nonces need to be stored:

\begin{lstlisting}
struct DHBalance {
	uint256 encryptedBalance;
    uint256[][2] sendersPublicKeys;
	uint256[] encryptionNonces;
}
\end{lstlisting}

\par Because both balance types have to be provided in the form of the pending and actual balances, the final balance storage looks like this:

\begin{lstlisting}
struct Balance {
	Commitment elGamalCommitmentPending;
    Commitment elGamalCommitmentActual;
	DHBalance dhEncryptedPending;
    DHBalance dhEncryptedActual;
}

mapping(uint256 => Balance) internal balances;
\end{lstlisting}

\par Note that internal balance accounting is performed using the x-coordinate of the user’s public key as the key in the balances mapping.

\subsection{Functions}
\subsubsection{Depositing to cWETH}
For the correct confidential account creation, the deposit function is to be provided:

\begin{lstlisting}
function deposit(
    uint256[2] publicKey,
    bytes calldata amountCommitmentData,
    bytes calldata balanceEncryptionData,
    bytes calldata proofData
) public payable;
\end{lstlisting}

\par Alongside a regular ZK proof, the proofData is required to make sure that the user actually owns the private key for the provided public key.
\par The amountCommitmentData is further decoded as follows:

\begin{lstlisting}
(uint256[2] commitmentC, uint256[2] commitmentD) =  
    abi.decode(amountCommitmentData, (uint256[2], uint256[2]));
\end{lstlisting}

\par The balanceEncryptionData is decoded as follows:

\begin{lstlisting}
(uint256 encryptedBalance, uint256 encryptionNonce) =   
    abi.decode(balanceEncryptionData, (uint256, uint256));
\end{lstlisting}

\subsubsection{Confidential transfer}

Having a clear understanding of the commitment and encryption schemes, the transfer function can be specified as follows:

\begin{lstlisting}
function transfer(
    address receiver,
    bytes calldata amountCommitmentData,
    bytes calldata amountEncryptionData,
    bytes calldata proofData
) external;
\end{lstlisting}

\par The amountCommitmentData is further decoded as follows:

\begin{lstlisting}
(
  uint256[2] senderCommitmentC,
  uint256[2] senderCommitmentD,
  uint256[2] receiverCommitmentC,
  uint256[2] receiverCommitmentD
) = abi.decode(
    amountCommitmentData, 
    (uint256[2], uint256[2], uint256[2], uint256[2])
  );
\end{lstlisting}

\par The amountEncryptionData is decoded as follows:

\begin{lstlisting}
(
  uint256 newEncryptedBalance,
  uint256 senderEncryptionNonce,
  uint256 receiverEncryptedAmount,
  uint256 receiverEncryptionNonce
) = abi.decode(
    amountEncryptionData,
    (uint256, uint256, uint256, uint256, uint256)
  );
\end{lstlisting}

\subsubsection{Withdrawing cWETH}
For unwrapping cWETH tokens, the withdraw function is to be provided:

\begin{lstlisting}
function withdraw(
	uint256 amount,
    bytes calldata amountCommitmentData,
    bytes calldata balanceEncryptionData,
    bytes calldata proofData
) external;
\end{lstlisting}

\par The amountCommitmentData is further decoded as follows:

\begin{lstlisting}
(uint256[2] commitmentC, uint256[2] commitmentD) =      
    abi.decode(amountCommitmentData, (uint256[2], uint256[2]));
\end{lstlisting}

\par The balanceEncryptionData is decoded as follows:

\begin{lstlisting}
(uint256 newEncryptedBalance, uint256 encryptionNonce) =   
    abi.decode(balanceEncryptionData, (uint256, uint256));
\end{lstlisting}

\section{Circom Circuits}
Each parameter described in this proposal is designed to be verifiable on-chain and compatible with zero knowledge circuits.

\subsection{Deposit to cWETH circuit}
The list of circuit signals for the deposit into the cWETH proof is the following:

\vspace{12pt}
Public signals:

\begin{itemize}
    \item User public key;
    \item Deposit amount;
    \item ElGamal commitment of the user's balance;
    \item ElGamal commitment of the deposit amount;
    \item User balance after the deposit encrypted with the user's public key-based DH key;
    \item Encryption random nonce.
\end{itemize}

Private signals:

\begin{itemize}
    \item User private key;
    \item Random nonce used in the ElGamal commitment.
\end{itemize}

Operating these signals, the circuit must have the following constraints:

\begin{enumerate}
  \item The provided private key is indeed the private key of the provided public key.
  \item The provided deposit amount is proven to be the one committed using the ElGamal-based commitment.
  \item The user balance after the deposit is proven to be the one encrypted with the DH shared key.
\end{enumerate}

\subsection{Confidential transfer circuit}

The list of circuit signals for the confidential transfer proof is the following:

\vspace{12pt}
Public signals:

\begin{itemize}
    \item Sender’s public key;
    \item Receiver’s public key;
    \item ElGamal commitment of the sender’s balance;
    \item ElGamal commitment of the transfer amount based on the sender’s public key;
    \item ElGamal commitment of the transfer amount based on the receiver’s public key;
    \item New sender’s balance encrypted with the sender’s public key-based DH shared key;
    \item Random nonce used in the sender’s new balance encryption;
    \item Transfer amount encrypted with the receiver's public key-based DH shared key;
    \item Random nonce used during the receiver’s transfer amount encryption.
\end{itemize}

Private signals:

\begin{itemize}
    \item Sender’s private key;
    \item Sender’s balance;
    \item Transfer amount;
    \item Random nonce used in the sender’s ElGamal commitment;
    \item Random nonce used in the receiver’s ElGamal commitment.
\end{itemize}

Operating these signals, the circuit must have the following constraints:

\begin{enumerate}
  \item The provided private key is indeed the private key of the provided sender’s public key.
  \item The provided sender’s balance is proven to be the one committed using the ElGamal commitment and to be greater than or equal to the transfer amount.
  \item The sender’s ElGamal commitment of the transfer amount was generated correctly.
  \item The receiver’s ElGamal commitment of the transfer amount was generated correctly.
  \item The new sender’s balance was correctly encrypted with the sender’s public key-based DH shared key.
  \item The transfer amount was correctly encrypted with the receiver’s public key-based DH shared key.
\end{enumerate}

\subsection{Withdraw from cWETH circuit}

The list of circuit signals for withdrawing cWETH proof is the following:

\vspace{12pt}
Public signals:

\begin{itemize}
    \item User public key;
    \item Receiver address;
    \item Withdrawal amount;
    \item ElGamal commitment of the user's balance;
    \item ElGamal commitment of the withdrawal amount based on the user's public key;
    \item Balance after the withdrawal encrypted with the user's public key-based DH shared key;
    \item Random nonce used during the new balance encryption.
\end{itemize}

Private signals:

\begin{itemize}
    \item User private key;
    \item User balance;
    \item Random nonce used in the ElGamal commitment.
\end{itemize}

Operating these signals, the circuit must have the following constraints:

\begin{enumerate}
  \item The provided private key is indeed the private key of the provided public key.
  \item The provided sender’s balance is proven to be the one committed using the ElGamal commitment and to be greater than or equal to the withdrawal amount.
  \item The ElGamal commitment of the withdrawal amount was generated correctly.
  \item The new balance (after the withdrawal) was correctly encrypted with the user's public key-based DH shared key.
\end{enumerate}

\printbibliography

\end{document}